\documentclass[aps,prd,showpacs,superscriptaddress]{revtex4}
\usepackage{graphicx}
\usepackage{bm}
\begin{document}
\title{Third-order perturbative solutions in the Lagrangian
perturbation theory with pressure II:
Effect of the transverse modes}
\author{Takayuki Tatekawa}
\email{tatekawa@gravity.phys.waseda.ac.jp}
\affiliation{Department of Physics, Waseda University,
3-4-1 Okubo, Shinjuku, Tokyo 169-8555, Japan}
\affiliation{Advanced Research Institute for Science
and Engineering, Waseda University,
3-4-1 Okubo, Shinjuku, Tokyo 169-8555, Japan}
\affiliation{Department of Physics, Ochanomizu University,
2-1-1 Ohtuka, Bunkyo, Tokyo 112-8610, Japan}
\date{\today}

\begin{abstract}
Lagrangian perturbation theory for cosmological fluid describes
structure formation in the quasi-nonlinear stage well.
In a previous paper, we presented
a third-order perturbative equation for Lagrangian perturbation
with pressure. There we considered
only the longitudinal modes for the first-order perturbation.
In this paper, we generalize the perturbation, i.e., we
consider both the longitudinal and the transverse modes
for the first-order perturbation. Then we derive
third-order perturbative equations and solutions.
\end{abstract}


\maketitle

\section{Introduction}\label{sec:intro}

The structure formation scenario based on gravitational instability
has been studied for a long time. The Lagrangian perturbative
method for the cosmological fluid describes the nonlinear
evolution of density fluctuation rather well. 
Zel'dovich~\cite{zel} proposed
a linear Lagrangian approximation for dust fluid.
This approximation is called the Zel'dovich approximation
(ZA)~\cite{zel,Arnold82,Shandarin89,buchert89,coles,saco,Jones04,Tatekawa05B,Paddy05}.
ZA describes the evolution of density fluctuation better than
the Eulerian approximation~\cite{Munshi94,Sahni96,yoshisato}.
Once ZA established, the second- and the third-order perturbative
solutions for dust fluid were derived
\cite{Barrow93,Bouchet92,Bouchet95,Buchert92,Buchert93,Buchert94,Catelan95,Sasaki98}.

Recently the effect of the pressure in the cosmological
fluid has been considered.
Buchert and Dom\'{\i}nguez~\cite{budo} showed that
when the velocity dispersion is regarded
as small and isotropic it produces effective ``pressure'' or
viscosity terms. Furthermore, they posited the
relation between mass density $\rho$ and pressure $P$, i.e.,
an ``equation of state.'' Recently,
Buchert and Dom\'{\i}nguez~\cite{Buchert05} discussed
the origin of ``adhesion'' in the structure formation scenario.
They considered two cases, velocity dispersion and
coarse-graining. They named these models
the Euler-Jeans-Newton (EJN) model and the Small-Size
Expansion (SSE) model, respectively.

In this paper, we treat only the EJN model.
Following our previous paper~\cite{Tatekawa05A},
we call this model the ``pressure model.''
Adler and Buchert~\cite{Adler99} have formulated the
Lagrangian perturbation theory for a barotropic fluid.
Morita and Tatekawa~\cite{Morita01} and Tatekawa
\textit{et al.}~\cite{Tatekawa02}
solved the Lagrangian perturbation equations for a polytropic fluid
up to the second order. In the aforementioned paper~\cite{Tatekawa05A},
we solved third-order perturbation equations
in the simple case.

In this paper, we notice the effect of the primordial vorticity.
In dust models, Buchert~\cite{Buchert92} investigated the behavior
of the vorticity with the first-order Lagrangina perturbative
equations. He showed that the primordial vorticity is amplified
in proportion to the enhancement of the density fluctuations
by deriving the relation between the vorticity and the density
fluctuations. Barrow and Saich~\cite{Barrow93} showed the effect
of the vorticity in gravitational collapse in an expanding Universe.
Sasaki and Kasai~\cite{Sasaki98} derived the solutions for
the perturbations of both the longitudinal and the transverse modes
up to third-order in E-dS Universe model. Then they showed how
the vorticity affects the evolution of the density fluctuation.

In the pressure model, although we showed the perturbative
equations for a polytropic fluid up to the second order
\cite{Morita01},
we ignored the effect of the first-order perturbation
in the transverse mode. As showed in the past paper
\cite{Sasaki98}, the effect of the vorticity becomes important
in nonlinear evolution of the density fluctuation in the dust model.
In the pressure model, the effect of the vorticity would
becomes important. Because the evolution of the first-order
perturbation in the transverse mode is independent of the
scale of the fluctuation, the effect especially appears in
small scale where the longitudinal mode shows decaying
oscillation.
Although we presented the third-order perturbative equation
for the pressure model~\cite{Tatekawa05A}, we ignored 
the effect of the primordial vorticity.
The transverse modes
do not have a growing solution in the first order, we could neglect
the transverse modes for the linear approximation.
However, the combination of longitudinal and transverse
modes do grow in third-order perturbation. Therefore
when we observe the transverse mode in the initial condition, we need
to estimate the effect of the transverse mode.
Then we derive the perturbative solution
for the simplest case, i.e., where the background is given by 
the Einstein-de Sitter (E-dS) Universe model and the polytropic index
$\gamma=4/3$. 

This paper is organized as follows.
In Sec.~\ref{sec:basic}, we present a Lagrangian description of
basic equations for cosmological fluids.
In Sec.~\ref{sec:Perturbation}, we show perturbative equations
and derive perturbative solutions for the pressure model
up to the third order.
In Sec.~\ref{subsec:1st-eq}, we show the perturbative equations
and solutions for generic polytropic fluid in the E-dS Universe model.
In Sec.~\ref{subsec:2nd-eq} and Sec.~\ref{subsec:3rd-eq},
we show the perturbative equations
for second- and third-order perturbation, respectively.
In Sec.~\ref{sec:Solution}, we derive perturbative solutions.
In general, it is extremely hard to solve higher-order perturbative
equations. Therefore we show generic solutions for first-order
perturbation only. In Sec.~\ref{subsec:2nd-sol} and
Sec.~\ref{subsec:3rd-sol}, we derive the perturbative solutions
for second- and third-order perturbation, respectively.
Here we derive the solutions only for the simplest case
 (E-dS Universe model, $\gamma=4/3$).
In Sec.~\ref{sec:summary}, we summarize our conclusions.

\section{Basic equations}\label{sec:basic}

We briefly introduce the Lagrangian description for cosmological
fluid. The basic equations for cosmological fluid were showed in
the previous paper~\cite{Tatekawa05A}. Here we show only the evolution
equation for Lagrangian perturbation.

In Newtonian cosmology,
to introduce cosmic expansion, we adopt the coordinate transformation
from physical coordinates to comoving coordinates.
\begin{equation} \label{eqn:comoving-coordinate}
\bm{x} = \frac{\bm{r}}{a(t)}, a(t):\mbox{scale factor} \,,
\end{equation}
where $\bm{r}$ and $\bm{x}$ are physical coordinates and
comoving coordinates, respectively.
The scale factor satisfies the Friedmann equations.

In the Lagrangian perturbation theory, instead of density fluctuation
the displacement from homogeneous distribution is regarded
as a perturbative quantity.
\begin{equation} \label{eqn:x=q+s}
\bm{r} = a \bm{x} = a ( \bm{q} + \bm{s} (\bm{q},t) ) \,,
\end{equation}
where $\bm{x}$ and $\bm{q}$ are the comoving Eulerian coordinates
and the Lagrangian coordinates, respectively. $\bm{s}$ is
the displacement vector. In the Lagrangian model, the displacement
vector is regarded as a perturbative quantity.
From Eq.~(\ref{eqn:x=q+s}), we can solve the continuous equation
exactly. Then the density
fluctuation is given in the formally exact form.
\begin{equation}
\delta = 1 - J^{-1}, ~~ J \equiv \det \left (
 \frac{\partial x_i}{\partial q_j} \right ) \,.
\end{equation}
$J$ means the Jacobian of the coordinate transformation
from Eulerian $\bm{x}$ to Lagrangian $\bm{q}$.
Therefore, when we derive the solutions for $\bm{s}$, we can know
the evolution of the density fluctuation. 

The peculiar velocity is given by
\begin{equation}
\bm{v}=a \dot{\bm{s}} \label{eqn:L-velocity} \,.
\end{equation}
Then we introduce the Lagrangian time derivative:
\begin{equation}
\frac{\rm d}{{\rm d} t} \equiv \frac{\partial}{\partial t}
 + \frac{1}{a} \bm{v} \cdot \nabla_x \,. \label{eqn:dt-L}
\end{equation}
Taking divergence and curl of Euler's equation, we obtain
the evolution equations for the Lagrangian displacement:
\begin{eqnarray}
\nabla_x
 \cdot \left (\ddot{\bm{s}}
  + 2 \frac{\dot{a}}{a} \dot{\bm{s}}
  - \frac{\kappa \gamma \rho_b^{\gamma-1}}{a^2}
 J^{-\gamma} \nabla_x J \right )
&=& -4 \pi G \rho_b (J^{-1} -1) \,, \label{eqn:L-longi-eqn} \\
\nabla_x \times \left (\ddot{\bm{s}} + 2 \frac{\dot{a}}{a}
 \dot{\bm{s}} \right )
&=& 0 \,, \label{eqn:L-trans-eqn}
\end{eqnarray}
where $(\dot{})$ means the Lagrangian time derivative (Eq.~(\ref{eqn:dt-L})).
To solve the perturbative equations,
we decompose the Lagrangian perturbation to the longitudinal
and transverse modes:
\begin{eqnarray}
\bm{s} &=& \nabla S + \bm{s}^T \,, \\
\nabla \cdot \bm{s}^T &=& 0 \,,
\end{eqnarray}
where $\nabla$ means the Lagrangian spacial derivative.

Here we expand the Jacobian:
\begin{eqnarray}
J &=& 1 + s_{i,i} + \frac{1}{2} \left ( s_{i,i} s_{j,j}
 - s_{i,j} s_{j,i} \right ) + \det \left (s_{i,j} \right )
 \nonumber \\
  &=& 1 + \nabla^2 S + \frac{1}{2} \left \{ (\nabla^2 S)^2
  - S_{,ij} S_{,ji} - s^T_{i,j} s^T_{j,i} 
  - 2 S_{,ij} s^T_{j,i} \right \} \nonumber \\
  &&  
  + \det \left (S_{,ij} + s^T_{i,j} \right ) \label{eqn:Jacobian}
\,.
\end{eqnarray}
Because Eqs.~(\ref{eqn:L-longi-eqn}) and (\ref{eqn:L-trans-eqn})
include the Eulerian spacial derivative, we change to the Lagrangian
spacial derivative.
\begin{eqnarray*}
\frac{\partial}{\partial x_i} &=& \frac{\partial}{\partial q_i}
 - s_{j,i} \frac{\partial}{\partial x_j} \\
 &=& \frac{\partial}{\partial q_i}
 - s_{j,i} \frac{\partial}{\partial q_j}
 + s_{j,i} s_{k,j} \frac{\partial}{\partial x_k} \\
 &=& \frac{\partial}{\partial q_i}
 - s_{j,i} \frac{\partial}{\partial q_j}
 + s_{j,i} s_{k,j} \frac{\partial}{\partial q_k} + \cdots \,.
\end{eqnarray*}

\section{The Lagrangian perturbative equations}
\label{sec:Perturbation}

\subsection{The first-order perturbations}
\label{subsec:1st-eq}

From Eqs.~(\ref{eqn:L-longi-eqn}) and (\ref{eqn:L-trans-eqn}),
we obtain the first-order perturbative equations:
\begin{eqnarray}
\nabla^2 \left ( \ddot{S}^{(1)} + 2 \frac{\dot{a}}{a}
 \dot{S}^{(1)} - 4\pi G \rho_b S^{(1)}
  - \frac{\kappa \gamma \rho_b^{\gamma-1}}{a^2} \nabla^2 S^{(1)}
 \right ) &=& 0 \,, \\
\nabla \times \left ( \ddot{\bm{s}}^{T (1)}
 + 2 \frac{\dot{a}}{a} \dot{\bm{s}}^{T (1)}
\right  ) &=& {\bm{0}} \,.
\end{eqnarray}
The first-order solutions for the longitudinal mode depend on
spacial scale. Therefore the solutions are described with a
Lagrangian wavenumber. In this paper, we discuss only perturbative
solutions in the E-dS Universe model~\cite{Morita01, Tatekawa02,Weinberg72}.
\begin{eqnarray}
\widehat{S}^{(1)} (\bm{K}, t) &=& C^+ (\bm{K}) D^+ (\bm{K}, t)
 + C^- (\bm{K}) D^- (\bm{K}, t) \,, \\
D^{\pm} (\bm{K}, t) &=& 
\left \{
\begin{array}{lcl}
t^{-1/6} {\cal J}_{\pm \nu} (A | \bm{K} | t^{-\gamma+4/3})
& \mbox{for} & \gamma \ne \frac{4}{3} \,, \\
t^{-1/6 \pm \sqrt{25/36 - B|\bm{K}|^2}}
& \mbox{for} & \gamma  = \frac{4}{3} \,, \\
\end{array}
\right. \label{eqn:1st-L-sol}
\end{eqnarray}
\[
A \equiv \frac{3 \sqrt{\kappa \gamma} (6 \pi G)^{(1-\gamma)/2}}{|4-3\gamma|},
~ B \equiv \frac{4}{3} \kappa (6 \pi G)^{-1/3} \,,
\]
where ${\cal J}$ means Bessel function. The coefficient $\nu$ is given
by $\nu=5/(8-6 \gamma)$.
$C^{\pm} (\bm{K})$ is given by the initial condition.

For the transverse mode, the solutions are same as for dust model:
\begin{equation}
{\bf s}^{T (1)} = C^{T+} + C^{T-} t^{-1/3} \,.
\end{equation}
The transverse mode does not have a growing solution. Therefore,
when we consider only large-scale fluctuation,
the longitudinal mode dominates during evolution. 

According to a comparison of the first-order perturbation
with a numerical simulation~\cite{Tatekawa04B},
the first-order perturbation seems rather good until the quasi-nonlinear
regime ($\delta \sim 1$). After that, we need to consider higher-order
perturbation.

\subsection{The second-order perturbative equations}
\label{subsec:2nd-eq}

From Eqs.~(\ref{eqn:L-longi-eqn}) and (\ref{eqn:L-trans-eqn}),
we obtain the second-order perturbative equations
\cite{Morita01,Weinberg72}.
For the longitudinal mode, the equation becomes
\begin{equation} \label{eqn:Lagrange-2ndL}
\nabla^2 \left ( \ddot{S}^{(2)} + 2 \frac{\dot{a}}{a}
 \dot{S}^{(2)} - 4\pi G \rho_b S^{(2)}
  - \frac{\kappa \gamma \rho_b^{\gamma-1}}{a^2} \nabla^2 S^{(2)}
 \right ) = Q^{L (2)} \,, 
\end{equation}
\begin{eqnarray}
Q^{L (2)} &=& 2 \pi G \rho_b \left [ S^{(1)}_{,ij} S^{(1)}_{,ij}
 - \left ( \nabla^2 S^{(1)} \right )^2 \right ] \nonumber \\
 && - \frac{\kappa \gamma \rho_b^{\gamma-1}}{a^2}
 \left \{ \gamma \nabla^2 S^{(1)}_{,i} \nabla^2 S^{(1)}_{,i} 
 + (\gamma-1) \nabla^2 \nabla^2 S^{(1)} \nabla^2 S^{(1)}
 + S^{(1)}_{,ijk} S^{(1)}_{,ijk} + 2 S^{(1)}_{,ij}
  \nabla^2 S^{(1)}_{,ij} \right \} \nonumber \\
 && - \frac{\kappa \gamma \rho_b^{\gamma-1}}{a^2}
 \left \{ S^{(1)}_{,ij} \nabla^2 s^{T (1)}_{i,j}
 + 2 S^{(1)}_{,ijk} s^{T (1)}_{i, jk}
 + \nabla^2 S^{(1)}_{,i} \nabla^2 s^{T (1)}_i 
 + 2 \nabla^2 S^{(1)}_{,ij} s^{T (1)}_{i,j} \right \} \nonumber \\
 && - 2 \pi G \rho_b s^{T (1)}_{i,j} s^{T (1)}_{j,i}
 - \frac{\kappa \gamma \rho_b^{\gamma-1}}{a^2}
 \left \{ s^{T (1)}_{i,j} \nabla^2 s^{T (1)}_{j,i}
 + s^{T (1)}_{i,jk} s^{T (1)}_{j,ik} \right \} \,.
\end{eqnarray}
For the transverse mode, after some arrangement, we can describe
as follows:
\begin{equation}
\nabla^2 \left ( \ddot{s}^{T (2)}_{i}
 + 2 \frac{\dot{a}}{a} \dot{s}^{T (2)}_{i} \right ) 
= Q_i^{T (2)} \,,
\end{equation}
\begin{eqnarray}
Q_i^{T (2)} & = & -4 \pi G \rho_b 
 \left \{ s^{T (1)}_{j,ik} S^{(1)}_{,jk}
 + s^{T (1)}_{j,i} \nabla^2 S^{(1)}_{,j}
 - \nabla^2 s^{T (1)}_{j} S^{(1)}_{,ij} - s^{T (1)}_{j,k} S^{(1)}_{,ijk}
 \right \} \nonumber \\
 && + \frac{\kappa \gamma \rho_b^{\gamma-1}}{a^2}
 \left \{ S^{(1)}_{,ijk} \nabla^2 S^{(1)}_{,jk} + S^{(1)}_{,ij}
 \nabla^2 \nabla^2 S^{(1)}_{,j}
 - \nabla^2 S^{(1)}_{,j}
 \nabla^2 S^{(1)}_{,ij} - S^{(1)}_{,jk} \nabla^2 S^{(1)}_{,ijk}
 \right \} \nonumber \\
 && + \frac{\kappa \gamma \rho_b^{\gamma-1}}{a^2}
 \left ( s^{T (1)}_{j,ik} \nabla^2 S^{(1)}_{,jk} + s^{T (1)}_{j,i}
 \nabla^2 \nabla^2 S^{(1)}_{,j}
 - \nabla^2 s^{T (1)}_{j}
 \nabla^2 S^{(1)}_{,ij} - s^{T (1)}_{j,k} \nabla^2 S^{(1)}_{,ijk}
 \right ) \,.
\end{eqnarray}
Here we can note second-order transverse mode solutions.
In the pressure model, even if we consider
only the longitudinal mode for the first order,
the second-order perturbation for the transverse mode
appears. In dust model, it does not appear.
Therefore, when we derive the third-order perturbative
solutions, we must consider the second-order
transverse mode.
\subsection{The third-order perturbative equations} \label{subsec:3rd-eq}

In the previous paper, we show the third-order perturbative equation
in which only the longitudinal modes for the first-order
perturbation considered~\cite{Tatekawa05A}.
The third-order perturbative equation becomes very complicated.
Here we introduce scalar quantities generated by 
the Lagrangian perturbations.
\begin{eqnarray}
\mu_1 (\Psi) & \equiv & \nabla^2 \Psi \,, 
 \label{eqn:mu1} \\
\mu_2 ({\bm A}, {\bm B}) & \equiv & \frac{1}{2}
 \left ( (\nabla \cdot {\bm A}) (\nabla \cdot {\bm B}) 
 - A_{i, j} B_{j, i} \right ) \,, \label{eqn:mu2AB} \\
\mu_2 ({\bm A}) & \equiv & \mu_2 ({\bm A}, {\bm A}) \,, 
 \label{eqn:mu2} \\
\mu_3 ({\bm A}) & \equiv & \mbox{det} \left (A_{i, j} \right ) \,,
 \label{eqn:mu3}
\end{eqnarray}
where ${\bm A}$ and ${\bm B}$ are vector quantities.
$\Psi$ is a scalar quantity. Using these quantities,
the Jacobian (Eq.~(\ref{eqn:Jacobian})) is written as
\begin{equation}
J = \mu_1 (S) + \mu_2 (S_{,i}+s^T_i) + \mu_3 (S_{,i} + s^T_i)
\,.
\end{equation}

First, we show the longitudinal mode equation. As in
the second-order perturbative equation, we separate the
terms of the third-order perturbation from the others. Then
the terms consisting of the first- and the second-order
perturbations are collected to the source term $Q^{L (3)}$.
\begin{equation}
\nabla^2 \left ( \ddot{S}^{(3)} + 2 \frac{\dot{a}}{a}
 \dot{S}^{(3)} - 4\pi G \rho_b S^{(3)}
  - \frac{\kappa \gamma \rho_b^{\gamma-1}}{a^2} \nabla^2 S^{(3)}
 \right ) = Q^{L (3)} \,. \label{eqn:3rd-L-eq}
\end{equation}
We consider the source term. Using
Eqs.~(\ref{eqn:mu1})-(\ref{eqn:mu3}), the terms are written as
follows:
\begin{eqnarray}
Q^{L (3)} &=& 4 \pi G \rho_b \left [ \mu_1 (S^{(1)})^3
 + \mu_2 (S^{(1)}_{,i} + s^{T (1)}_i, S^{(2)}_{,i} + s^{T (2)}_i)
 + \mu_3 (S^{(1)}_{,i} + s^{T (1)}_i) \right . \nonumber \\
&& ~~~~~~~~~ \left . - (S^{(1)}_{,ij} + s^{T (1)}_{j,i})
 (S^{(1)}_{,jk} + s^{T (1)}_{k,j}) S^{(1)}_{,ik}
 + (S^{(2)}_{,ij} + s^{T (2)}_{j,i} ) S^{(1)}_{,ij}
 \right ] \nonumber \\
&& +  (S^{(1)}_{,ij} + s^{T (1)}_{j,i})
 \left (\ddot{S}^{(2)}_{,ij} + 2 \frac{\dot{a}}{a} \dot{S}^{(2)}_{,ij}
 + \ddot{s}^{T (2)}_{i,j} + 2 \frac{\dot{a}}{a} \dot{s}^{T (2)}_{i,j}
 \right ) \nonumber \\
&& + \frac{\kappa \gamma \rho_b^{\gamma-1}}{a^2}
 \left [ \mu_2 ( S^{(1)}_{,i} + s^{T (1)}_{i},
   S^{(2)}_{,i} + s^{T (2)}_i )
 + \mu_3 (S^{(1)}_{,i}) \right . \nonumber \\
&& ~~~~~~~~~~ \left . - (S^{(1)}_{,jk} + s^{T (1)}_{k,j})
  \left \{ \mu_1 (S^{(2)}) 
   + \mu_2 ( S^{(1)}_{,i} + s^{T (1)}_{i} ) \right \}_{,jk}
 \right . 
 \nonumber \\
&& ~~~~~~~~~~ \left . - \left \{ (S^{(1)}_{,jk} + s^{T (1)}_{k,j})
  \left ( \mu_1 (S^{(2)}) 
   + \mu_2 ( S^{(1)}_{,i} + s^{T (1)}_{i} ) \right )_{,k} 
  \right \}_{,j} \right . 
 \nonumber \\
&& ~~~~~~~~~~ \left . - \left \{ \gamma \mu_1 (S^{(1)}) 
 \left ( \mu_1 (S^{(2)}) +
  \mu_2 (S^{(1)}_{,i} + s^{T (1)}_i) \right )_{,j}
  \right \}_{,j} \right .
 \nonumber \\
&& ~~~~~~~~~~ \left . + \left \{ 
  (S^{(1)}_{,jk} + s^{T (1)}_{k,j} ) (S^{(1)}_{,kl} + s^{T (1)}_{l,k} )
 \nabla^2 S^{(1)}_{,l} - (S^{(2)}_{,jk} + s^{T (2)}_{k,j} )
 \nabla^2 S^{(1)}_{,k} \right \}_{,j} 
\right . \nonumber \\
&& ~~~~~~~~~~ \left . - \left \{
  \gamma \left ( \mu_1 (S^{(2)})
  + \mu_2 (S^{(1)}_{,i} + s^{T (1)}_i)
  \right ) \nabla^2 S^{(1)}_{,j}
 - \frac{\gamma (\gamma+1)}{2} \mu_1 (S^{(1)})^2
 \mu_1  (S^{(1)})_{,j} \right \}_{,j}
 \right . \nonumber \\
&& ~~~~~~~~~~ \left . + \gamma \left \{ \mu_1 (S^{(1)})
 (S^{(1)}_{,jk} + s^{T (1)}_{k,j}) \mu_1 (S^{(1)})_{,k}
 \right \}_{,j}
 \right . \nonumber \\
&& ~~~~~~~~~~ \left . + (S^{(1)}_{,jk} + s^{T (1)}_{k,j} )
 \left \{ (S^{(1)}_{,jl} + s^{T (1)}_{l,j} )
 \mu_1 (S^{(1)})_{,l} \right \}_{,k}
 \right . \nonumber \\
&& ~~~~~~~~~~ \left . + \gamma 
 (S^{(1)}_{,jk} + s^{T (1)}_{k,j} )
 \left \{ \mu_1 (S^{(1)})
 \mu_1 (S^{(1)})_{,j} \right \}_{,k}
 \right ] \,.
 \label{eqn:3rd-L}
\end{eqnarray}
The transverse mode also seems complicated. However, if we
neglect the first-order transverse mode, the evolution equation
is described as
\begin{equation}
-\nabla^2 \left ( \ddot{s}^{T (3)}_i + 2 \frac{\dot{a}}{a}
 \ddot{s}^{T (3)}_i \right ) =Q_i^{T (3)} \,,
\end{equation}
\begin{eqnarray}
Q_i^{T (3)} &=& \left \{ \left ( S^{(1)}_{,ij} + s^{T (1)}_{j,i} \right )
 \left ( \ddot{S}^{(2)}_{,jk}
 + 2 \frac{\dot{a}}{a} \dot{S}^{(2)}_{,jk}
 + \ddot{s}^{T (2)}_{k,j}
 + 2 \frac{\dot{a}}{a} \dot{s}^{T (2)}_{k,j}
 \right ) \right \}_{,k}
 \nonumber \\
&& - \left \{ \left (S_{,jk}^{(1)} + s^{T (1)}_{j,k} \right )
 \left ( \ddot{S}^{(2)}_{,ij}
 + 2 \frac{\dot{a}}{a} \dot{S}^{(2)}_{,ij}
 + \ddot{s}^{T (2)}_{i,j}
 + 2 \frac{\dot{a}}{a} \dot{s}^{T (2)}_{i,j} \right ) \right \}_{,k}
 \nonumber \\
&&  + \left \{ \left ( S_{,ij}^{(2)} + s_{i,j}^{T (2)} \right )
 \left ( 4 \pi G \rho_b S^{(1)}_{,jk}
 + \frac{\kappa \gamma \rho_b^{\gamma-1}}{a^2} \nabla^2 S^{(1)}_{,jk}
 \right ) \right \}_{,k} \nonumber \\
&&  - \left \{ \left ( S_{,jk}^{(2)} + s_{j,k}^{T (2)} \right )
 \left ( 4 \pi G \rho_b S^{(1)}_{,ij}
 + \frac{\kappa \gamma \rho_b^{\gamma-1}}{a^2} \nabla^2 S^{(1)}_{,ij}
  \right )  \right \}_{,k} \nonumber \\
&& + \frac{\kappa \gamma \rho_b^{\gamma-1}}{a^2}
 \left ( \nabla^2 S_{,il}^{(1)} S_{,jk}^{(1)} S_{,jkl}^{(1)}
 + \nabla^2 S_{,ik}^{(1)} \nabla^2 S_{,j}^{(1)} S_{,jk}^{(1)}
 + \nabla^2 S_{,ikl}^{(1)} S_{,jk}^{(1)} S_{,jl}^{(1)} \right . \nonumber \\
&& ~~~~~~~~ \left . - S_{,ijk}^{(1)} S_{,jl}^{(1)} \nabla^2 S_{,kl}^{(1)}
 - S_{,ij}^{(1)} S_{,jkl}^{(1)} \nabla^2 S_{,kl}^{(1)}
 - S_{,ij}^{(1)} S_{,jk}^{(1)} \nabla^2 S_{,k}^{(1)}
 \right ) \nonumber \\ 
&& + 4 \pi G \rho_b \left [
 \nabla^2 s^{T (1)}_{k} S^{(1)}_{,ij} S^{(1)}_{,jk}
 -s^{T (1)}_{l,ij} S^{(1)}_{,kl} S^{(1)}_{,jk}
  \right . \nonumber \\
&& ~~~~~~~~ \left . + s^{T (1)}_{k,j} \left (
 \nabla^2 S^{(1)}_{,j} S^{(1)}_{,ik}
 - \nabla^2 S^{(1)}_{,k} S^{(1)}_{,ij}
 + S^{(1)}_{,jl} S^{(1)}_{,ikl}
 - S^{(1)}_{,ijl} S^{(1)}_{,kl}
 + S^{(1)}_{,il} S^{(1)}_{,jkl}
 + S^{(1)}_{,kl} S^{(1)}_{,ijl}
 \right )
 \right . \nonumber \\
&& ~~~~~~~~ \left . - s^{T (1)}_{j,i} \left (
 S^{(1)}_{,jkl} S^{(1)}_{,jk} + S^{(1)}_{,jk} \nabla^2 S^{(1)}_{,k}
 \right ) \right . \nonumber \\ 
&& ~~~~~~~~ \left . + s^{T (1)}_{k,jl}
 \left ( S^{(1)}_{,jl} S^{(1)}_{,ik} - S^{(1)}_{,il} S^{(1)}_{,jk}
 \right ) \right ]
 \nonumber \\
&& + \frac{\kappa \gamma \rho_b^{\gamma-1}}{a^2} \left [
 \nabla^2 s^{T (1)}_{k} S^{(1)}_{,jk} \nabla^2 S^{(1)}_{,ij}
 -s^{T (1)}_{l,ij} S^{(1)}_{,kl} \nabla^2 S^{(1)}_{,jk}
  \right . \nonumber \\
&& ~~~~~~~~ \left . + s^{T (1)}_{k,j} \left (
 \nabla^2 S^{(1)}_{,j} \nabla^2 S^{(1)}_{,ik}
 - \nabla^2 \nabla^2 S^{(1)}_{,k} S^{(1)}_{,ij}
 + S^{(1)}_{,jl} \nabla^2 S^{(1)}_{,ikl} \right . \right.
 \nonumber \\
&& ~~~~~~~~~~~~ \left. \left. 
 - S^{(1)}_{,ijl} \nabla^2 S^{(1)}_{,kl}
 + S^{(1)}_{,jkl} \nabla^2 S^{(1)}_{,il}
 + S^{(1)}_{,kl} \nabla^2 S^{(1)}_{,ijl}
 \right )
 \right . \nonumber \\
&& ~~~~~~~~ \left . - s^{T (1)}_{j,i} \left (
 S^{(1)}_{,jkl} \nabla^2 S^{(1)}_{,kl}
 + S^{(1)}_{,jk} \nabla^2 \nabla^2 S^{(1)}_{,k}
 \right ) \right . \nonumber \\ 
&& ~~~~~~~~ \left . + s^{T (1)}_{k,jl}
 \left ( S^{(1)}_{,jl} \nabla^2 S^{(1)}_{,ik}
 - S^{(1)}_{,il} \nabla^2 S^{(1)}_{,jk}
 \right ) \right ]
 \nonumber \\
&& + 4 \pi G \rho_b \left [
 -\left (s^{T (1)}_{l,ij} s^{T (1)}_{k,l}
 + s^{T (1)}_{l,i} s^{T (1)}_{k,jl} \right ) S^{(1)}_{,jk}
 - s^{T (1)}_{k,i} s^{T (1)}_{j,k} \nabla^2 S^{(1)}_{,j}
 \right . \nonumber \\
&& ~~~~~~~~ \left .
 + \left (\nabla^2 s^{T (1)}_{k} s^{T (1)}_{j,k} +
  s^{T (1)}_{l,k} s^{T (1)}_{j,kl} \right ) S^{(1)}_{,ij}
 + s^{T (1)}_{l,j} s^{T (1)}_{k,l} S^{(1)}_{,ijk}
 \right ] \nonumber \\
&& + \frac{\kappa \gamma \rho_b^{\gamma-1}}{a^2}
 \left [ - \left (s^{T (1)}_{l,ij} s^{T (1)}_{k,l}
 + s^{T (1)}_{l,i} s^{T (1)}_{k,jl} \right )
 \nabla^2 S^{(1)}_{,jk}
 - s^{T (1)}_{k,i} s^{T (1)}_{j,k} \nabla^2 \nabla^2 S^{(1)}_{,j}
 \right . \nonumber \\
&& ~~~~~~~~ \left .
 + \left (\nabla^2 s^{T (1)}_{k} s^{T (1)}_{j,k} +
  s^{T (1)}_{l,k} s^{T (1)}_{j,kl} \right )
  \nabla^2 S^{(1)}_{,ij}
 + s^{T (1)}_{l,j} s^{T (1)}_{k,l} \nabla^2 S^{(1)}_{,ijk}
 \right ]\,.
\end{eqnarray}
%

\section{Lagrangian perturbative solutions for higher-order
perturbation} \label{sec:Solution}

\subsection{The second-order perturbative solutions:
The simplest case} \label{subsec:2nd-sol}

The second-order solutions are formally written as follows:
\begin{eqnarray}
\widehat{S}^{(2)} &=& -\frac{1}{|\bm{K}|^2}
 \int^t {\rm d} t' G(\bm{K}, t, t')
 \widehat{Q}^{L (2)} (\bm{K}, t) \,, \\
\widehat{s}^{T (2)} &=& -\frac{1}{|\bm{K}|^2}
 \int^t {\rm d} t' G^T(t, t')
 \widehat{Q}^{T (2)} (\bm{K}, t) \,, \\
G^T (t, t') &=& 3 (t'- t^{-1/3} t'^{4/3} ) \,.
\end{eqnarray}
$G^L (\bm{K}, t, t')$ depends on the ``equation of state.''
If $\gamma \ne 4/3$
and $\nu = 5/(8-6\gamma)$ is not an integer, we have
\begin{eqnarray}
G^L (\bm{K}, t, t') &=& -\frac{\pi}{2 \sin \nu \pi}
 \left (-\gamma + \frac{4}{3} \right )^{-1} t^{-1/6} t'^{7/6}
 \left [ \mathcal{J}_{-\nu} \left (A |\bm{K}| t^{-\gamma+4/3} \right )
  \mathcal{J}_{\nu} \left (A |\bm{K}| t'^{-\gamma+4/3} \right )
 \right . \nonumber \\
 &&~~ \left. - \mathcal{J}_{\nu} \left (A |\bm{K}| t^{-\gamma+4/3} \right )
  \mathcal{J}_{-\nu} \left (A |\bm{K}| t'^{-\gamma+4/3} \right )
 \right ] \,,
\end{eqnarray}
and if $\gamma=4/3$,
\begin{eqnarray}
G^L (\bm{K}, t, t') &=& -\frac{1}{2 b(\bm{K})} t^{-1/6} t'^{7/6}
 \left (t^{-b(\bm{K})} t'^{b(\bm{K})} 
 - t^{b(\bm{K})} t'^{-b(\bm{K})} \right ) \,, \\
b(\bm{K}) & \equiv & \sqrt{\frac{25}{36} - B | \bm{K} |^2} \,.
\label{eqn:b-definition}
\end{eqnarray}

The second-order perturbative solutions for the case
of $\gamma=4/3$ in the E-dS universe model has already been derived
by Morita and Tatekawa~\cite{Morita01}. The solutions
are described by
\begin{eqnarray}
\hat{S}^{(2)} (\bm{K}) &=& 
 \int {\rm d} \bm{K}' ~\left ( \Xi_1^L (\bm{K}, \bm{K}')
 E_1^L (\bm{K}, \bm{K}', t) + \Xi_{2~i}^L (\bm{K}, \bm{K}')
 E_{2~i}^L (\bm{K}, \bm{K}', t) \right . \nonumber \\
&& ~~~~~~~~~~~~ \left . +\Xi_{3~ij}^L (\bm{K}, \bm{K}')
 E_{3~ij}^L (\bm{K}, \bm{K}', t)
\right ) \,, \\
\hat{s}_i^{T (2)} (\bm{K}) &=& 
 \int {\rm d} {\bf K}' ~\left ( \Xi^T_{1~i} (\bm{K}, \bm{K}')
 E_{1}^T (\bm{K}, \bm{K}', t) + \Xi^T_{2~ij} (\bm{K}, \bm{K}')
 E_{2~j}^T (\bm{K}, \bm{K}', t)
 \right ) \,,
\end{eqnarray}
\begin{eqnarray}
\Xi_1^L (\bm{K}, \bm{K}') &=& -\frac{1}{(2 \pi)^3} \frac{1}{|\bm{K}|^2}
\left [
 \frac{1}{3} \left \{ (\bm{K}' \cdot (\bm{K}-\bm{K}'))^2
 - |\bm{K}'|^2 |\bm{K}-\bm{K}'|^2 \right \} \right . \nonumber \\
&& ~~ \left . + B \left \{ \frac{4}{3}
  |\bm{K}'|^2 |\bm{K}-\bm{K}'|^2
  (\bm{K}' \cdot (\bm{K}-\bm{K}'))
  + \frac{1}{3} |\bm{K}'|^4 |\bm{K}-\bm{K}'|^2 
  \right . \right . \nonumber \\
&& ~~~~~~~~~~~~ \left . \left . 
  + \left ( (\bm{K}' \cdot (\bm{K}-\bm{K}'))^3
  + 2 |\bm{K}-\bm{K}'|^2 (\bm{K}' \cdot (\bm{K}-\bm{K}'))^2
  \right ) \right \}  \right ] \,, \\
\Xi_{2~i}^L (\bm{K}, \bm{K}') &=& -\frac{i B}{(2 \pi)^3} \frac{1}{|\bm{K}|^2}
\left [ |\bm{K}'|^2  (\bm{K}' \cdot (\bm{K}-\bm{K}'))
  + 2 (\bm{K}' \cdot (\bm{K}-\bm{K}'))^2 \right . \nonumber \\
&& ~~~~~~~~~~~~ \left .
  + |\bm{K}'|^2 |\bm{K}-\bm{K}'|^2 + 2 |\bm{K}'|^2 
 (\bm{K}' \cdot (\bm{K}-\bm{K}')) \right ] K'_i \,, \\
\Xi_{3~ij}^L (\bm{K}, \bm{K}') &=& -\frac{1}{(2 \pi)^3} \frac{1}{|\bm{K}|^2}
\left [ \frac{1}{3} (\bm{K}' \cdot (\bm{K}-\bm{K}'))
 - B \left \{ |\bm{K}-\bm{K}'|^2
 +  (\bm{K}' \cdot (\bm{K}-\bm{K}')) \right \}
 \right ] K'_i (K_j-K'_j) \,, \\
\Xi^T_{1~i} (\bm{K}, \bm{K}') &=& - \frac{i}{(2 \pi)^3} \frac{B}{|\bm{K}|^2}
 |\bm{K}-\bm{K}'|^2 (\bm{K}' \cdot (\bm{K}-\bm{K}'))
 \nonumber \\
&& ~~~~
 \left [ (\bm{K}' \cdot (\bm{K}-\bm{K}')) K'_i
  + |\bm{K}-\bm{K}'|^2 K'_i
  - |\bm{K}'|^2 (K_i - K'_i)
  - (\bm{K}' \cdot (\bm{K}-\bm{K}')) (K_i - K'_i)
 \right ] \,, \\
\Xi^T_{2~ij} (\bm{K}, \bm{K}') &=& \frac{1}{(2 \pi)^3}
 \left [ \frac{1}{3} \left \{ \left (
 (\bm{K}' \cdot (\bm{K}-\bm{K}')) + |\bm{K}-\bm{K}'|^2
  \right ) K'_i \right . \right . \nonumber \\
&& ~~~~~~~~ \left . \left .
  - \left ( |\bm{K}'|^2 + (\bm{K}' \cdot (\bm{K}-\bm{K}')) \right )
   (K_i-K'_i) \right \} (K_j - K'_j) \right . \nonumber \\
&& ~~~~ \left . 
 + B \left \{ \left ( 
 (\bm{K}' \cdot (\bm{K}-\bm{K}')) |\bm{K}-\bm{K}'|^2
  + |\bm{K}-\bm{K}'|^4 \right ) K'_i \right . \right . \nonumber \\
&& ~~~~~~~~ \left . \left .
 - \left ( |\bm{K}'|^2 |\bm{K}-\bm{K}'|^2
 - |\bm{K}-\bm{K}'|^2 (\bm{K}' \cdot (\bm{K}-\bm{K}'))
 \right ) (K_i - K'_i) \right \} (K_j - K'_j)
 \right ] \,.
\end{eqnarray}
\begin{eqnarray}
E^L_1 (\bm{K}, \bm{K}', t) & \propto & \frac{1}{2 b(\bm{K})}
 \sum_{(\oplus=\pm)} \sum_{(\otimes=\pm)} \left [
    \frac{C^{\oplus} (\bm{K}') C^{\otimes} (\bm{K}-\bm{K}')
     t^{-1/3 \oplus b(\bm{K}') \otimes b(\bm{K}-\bm{K}')}}
    {((\oplus b(\bm{K}') \otimes b(\bm{K}-\bm{K}')-\frac{1}{6})^2
     - b(\bm{K})^2)} \right ] \,, \\
E^L_{2~i} (\bm{K}, \bm{K}', t) & \propto & \frac{1}{2 b(\bm{K})}
 \sum_{(\oplus=\pm)} \sum_{(\otimes=\pm)}
    \frac{C^{T\oplus} (\bm{K}') C^{\otimes} (\bm{K}-\bm{K}')
  t^{-1/3 \oplus 1/6 \otimes b(\bm{K}')} }
  {\left (b(\bm{K}')-\frac{1}{6} \oplus \frac{1}{6}\right )^2
   - b(\bm{K})^2 } \,, \\
E^L_{3~ij} (\bm{K}, \bm{K}', t) & \propto & \frac{1}{2 b(\bm{K})}
  \left [ \frac{C^{T+} (\bm{K}') C^{T+} (\bm{K}-\bm{K}')}
  {\frac{1}{36} - b(\bm{K})^2}
 + \frac{C^{T+} (\bm{K}') C^{T-} (\bm{K}-\bm{K}')}
  {\frac{1}{36} - b(\bm{K})^2} t^{-1/3} \right . \nonumber \\
 && ~~~~~~~~ \left . + \frac{C^{T-} (\bm{K}') C^{T+} (\bm{K}-\bm{K}')}
  {\frac{1}{36} - b(\bm{K})^2} t^{-1/3}
  + \frac{C^{T-} (\bm{K}') C^{T-} (\bm{K}-\bm{K}')}
  {\frac{1}{4} - b(\bm{K})^2} t^{-2/3} \right ] \,, \\
E^T_1 (\bm{K}, \bm{K}', t) & \propto & 3 \sum_{(\oplus=\pm)}
     \sum_{(\otimes=\pm)} \left [
    \frac{C^{\oplus} (\bm{K}') C^{\otimes} (\bm{K}-\bm{K}')
     t^{-1/3 \oplus b(\bm{K}') \otimes b(\bm{K}-\bm{K}')}}
    {(\oplus b(\bm{K}') \otimes b(\bm{K}-\bm{K}')-\frac{1}{3})
     (\oplus b(\bm{K}') \otimes b(\bm{K}-\bm{K}'))} \right ] \,, \\
E^T_{2~j} (\bm{K}, \bm{K}', t) & \propto & \sum_{(\oplus=\pm)}
 \left [ \frac{C^{T\oplus} (\bm{K}') C^{+} (\bm{K}-\bm{K}')}
 {b(\bm{K})^2-\frac{1}{36}}
 t^{-1/6 \oplus b(\bm{K})} \right . \nonumber \\
 && ~~~~~~~~ \left . 
 + \frac{C^{T\oplus} (\bm{K}') C^{-} (\bm{K}-\bm{K}')}
 {\left (b(\bm{K}-\bm{K}') \oplus \left(-\frac{1}{2} \right ) \right )
  \left (b(\bm{K}-\bm{K}') \oplus \left(-\frac{1}{6} \right ) \right )}
 t^{-1/2 \oplus b(\bm{K}-\bm{K}')} \right ] \,.
\end{eqnarray}
where $\sum_{(\oplus=\pm)}$ means
\begin{equation}
\sum_{(\oplus=\pm)} (\alpha^{\oplus} \oplus \beta)
 \equiv (\alpha^+ + \beta) + (\alpha^- - \beta) \,.
\end{equation}
$E^L_2$ and $E^T_1$ appears only in the pressure model. If we ignore
the effect of the pressure, only $E^L_1, E^L_3$, and $E^T_2$ appear.
From these results, it is apparent that the pressure does affect
the evolution of the higher-order perturbation.

\subsection{The third-order perturbative solutions:
The simplest case} \label{subsec:3rd-sol}

As in the first- and the second-order solutions, the third-order
solutions are described with the Lagrangian wavenumber. Following the
method used in the second-order solutions,
the third-order solution is given by this integration:
\begin{eqnarray}
\widehat{S}^{(3)} &=& -\frac{1}{|\bm{K}|^2}
 \int_{t_0}^t {\rm d} t' G(\bm{K}, t, t')
 \widehat{Q}^{L (3)} (\bm{K}, t) \,, \\
\widehat{s}^{T (3)} &=& -\frac{1}{|\bm{K}|^2}
 \int_{t_0}^t {\rm d} t' G^T(t, t')
 \widehat{Q}^{T (3)} (\bm{K}, t) \,,
\end{eqnarray}
In the previous paper~\cite{Tatekawa05A}, we showed
the third-order perturbative solutions for the simplest
case, the case of $\gamma=4/3$ in the E-dS Universe model.
In this case, the contributions of the gravitational terms and
the pressure terms become identical:
\begin{equation}
4 \pi G \rho_b, \frac{\kappa \gamma \rho_b^{\gamma-1}}{a^2}
 \propto a^{-3} \,.
\end{equation}
In addition to these conditions, we considered only the longitudinal
modes in the first-order perturbation.
Here we consider both the longitudinal and the transverse
modes in the first-order perturbation. As we show in
Sec.~\ref{subsec:3rd-eq}, the third-order perturbative
equations become more complicated. To avoid complexity,
we estimate only time evolution in the third-order perturbative
solutions $F(t)$:
\begin{equation} \label{eqn:F-alpha}
F(t) \propto t^{\alpha} \,.
\end{equation}
The third-order perturbation is affected from the first-
and the second-order perturbation via source terms
$Q^L$ and $Q^T$:
\begin{eqnarray}
Q^{L (3)} &=& Q^{L (3)} \left (S^{(1)} \ast S^{(2)}, S^{(1)} \ast s^{T (2)},
 s^{T (1)} \ast S^{(2)}, s^{T (1)} \ast s^{T (2)}, \right . \nonumber \\
&& ~~~~ \left. S^{(1)} \ast S^{(1)} \ast S^{(1)},
 S^{(1)} \ast S^{(1)} \ast s^{T (1)},
 S^{(1)} \ast s^{T (1)} \ast s^{T (1)},
 s^{T(1)} \ast s^{T (1)} \ast s^{T (1)} \right ) \,, \\
Q^{T (3)} &=& Q^{T (3)} \left (S^{(1)} \ast S^{(2)}, S^{(1)} \ast s^{T (2)},
 s^{T (1)} \ast S^{(2)}, s^{T (1)} \ast s^{T (2)}, \right . \nonumber \\
&& ~~~~ \left. S^{(1)} \ast S^{(1)} \ast S^{(1)},
 S^{(1)} \ast S^{(1)} \ast s^{T (1)},
 S^{(1)} \ast s^{T (1)} \ast s^{T (1)} \right ) \,,
\end{eqnarray}
where $\ast$ means convolution.

In Sec.~\ref{subsec:1st-eq} and \ref{subsec:2nd-sol},
we showed the time dependence of the perturbative solutions.
Here we arrange and list again:
\begin{eqnarray}
D^{\pm} (\bm{K}) & \propto & t^{-1/6 \pm b(\bm{K})} \,, \\
D^{T\pm} & \propto & t^{-1/6 \pm 1/6} \,, \\
E_1^L (\bm{K}, \bm{K}') & \propto & t^{-1/3 \pm b(\bm{K}')
 \pm b(\bm{K}-\bm{K}')} \,, \\
E_2^L (\bm{K}, \bm{K}') & \propto & t^{-1/3 \pm 1/6
 \pm b(\bm{K}-\bm{K}')} \,, \\
E_3^L (\bm{K}, \bm{K}') & \propto & t^{-1/3 \pm 1/6
 \pm 1/6} \,, \\
E_1^T (\bm{K}, \bm{K}') & \propto & t^{-1/3 \pm b(\bm{K}')
 \pm b(\bm{K}-\bm{K}')} \,, \\
E_2^T (\bm{K}, \bm{K}') & \propto & t^{-1/3 \pm 1/6
 \pm b(\bm{K}-\bm{K}')} \,.
\end{eqnarray}
From Eq.~(\ref{eqn:b-definition}), if we take a limit of weak
pressure ($P \rightarrow 0$), $b(\bm{K})$ converges to
\begin{equation} \label{eqn:b-limit}
b(\bm{K}) \rightarrow \frac{5}{6} \,.
\end{equation}

Here we notice the relation between $E_1^L$ and $E_1^T$,
i.e., time evolution in both the longitudinal and the transverse
modes. 
Both of these terms are generated from $S^{(1)} \ast S^{(1)}$
in $Q^{L (2)}$ and $Q^{T (2)}$. Therefore if the convolution
in the source terms in both modes are identical, the 
evolutions will also be identical.

\begin{table}
\caption{\label{tab:F-L-alpha} The time dependence of the third-order
perturbative solutions in the longitudinal modes
(the case of $\gamma=4/3$ in the E-dS universe model).
The contribution means the time component in the source terms.
The indices of the time evolution $\alpha$ 
(Eq.~(\ref{eqn:F-alpha})) depend
on the convolution in the source terms $Q^{L (3)}$.
Here we define 
$b_1 \equiv b(\bm{K}''), b_2 \equiv b(\bm{K}'-\bm{K}''), 
 b_3 \equiv b(\bm{K}-\bm{K}'), b_4 \equiv b(\bm{K}-\bm{K}'')$.
``P'' denotes that the contribution does not appear
in the dust model.}
\begin{ruledtabular}
\begin{tabular}{lllc}
The convolution & The contribution & The indices $\alpha$ &\\ \hline
$S^{(1)} (\bm{K}'') \ast S^{(2)} (\bm{K}-\bm{K}'')$
& $D E^L_1$ & $-\frac{1}{2} \pm b_1 \pm b_2
\pm b_3$ & \\
& $D E^L_2$ & $-\frac{1}{2} \pm \frac{1}{6} \pm b_1
\pm b_4$ & P \\
& $D E^L_3$ & $-\frac{1}{2} \pm \frac{1}{6} \pm
 \frac{1}{6} \pm b_1$ & \\
$S^{(1)} (\bm{K}'') \ast s^{T (2)} (\bm{K}-\bm{K}'')$
& $D E^T_1$ & $-\frac{1}{2} \pm b_1 \pm b_2
\pm b_3$ & P \\
& $D E^T_2$ & $-\frac{1}{2} \pm \frac{1}{6} \pm b_1
\pm b_4$ & \\
$s^{T (1)} (\bm{K}'') \ast S^{(2)} (\bm{K}-\bm{K}'')$
& $D^T E^L_1$ & $-\frac{1}{2} \pm \frac{1}{6} \pm b_2
\pm b_3$ & \\
& $D^T E^L_2$ & $-\frac{1}{2} \pm \frac{1}{6} \pm \frac{1}{6}
\pm b_4$ & P \\
& $D^T E^L_3$ & $-\frac{1}{2} \pm \frac{1}{6} \pm \frac{1}{6}
\pm \frac{1}{6}$ & \\
$s^{T (1)} (\bm{K}'') \ast s^{T (2)} (\bm{K}-\bm{K}'')$
& $D^T E^T_1$ & $-\frac{1}{2} \pm \frac{1}{6} \pm b_2
\pm b_3$ & P \\
& $D^T E^T_2$ &  $-\frac{1}{2} \pm \frac{1}{6} \pm \frac{1}{6}
\pm b_4$ & \\
$S^{(1)} (\bm{K}'') \ast S^{(1)} (\bm{K}'-\bm{K}'')
 \ast S^{(1)} (\bm{K}-\bm{K}')$
& $D D D$ & $-\frac{1}{2} \pm b_1 \pm b_2
\pm b_3$ & \\
$s^{T (1)} (\bm{K}'') \ast S^{(1)} (\bm{K}'-\bm{K}'')
 \ast S^{(1)} (\bm{K}-\bm{K}')$
& $D^T D D$ & $-\frac{1}{2} \pm \frac{1}{6} \pm b_2
\pm b_3$ & \\
$s^{T (1)} (\bm{K}'') \ast s^{T (1)} (\bm{K}'-\bm{K}'')
 \ast S^{(1)} (\bm{K}-\bm{K}')$
& $D^T D^T D$ & $-\frac{1}{2} \pm \frac{1}{6} \pm \frac{1}{6}
\pm b_3$ & P \\
$s^{T (1)} (\bm{K}'') \ast s^{T (1)} (\bm{K}'-\bm{K}'')
 \ast s^{T (1)} (\bm{K}-\bm{K}')$
& $D^T D^T D^T$ & $-\frac{1}{2} \pm \frac{1}{6} \pm \frac{1}{6}
\pm \frac{1}{6}$ &
\end{tabular}
\end{ruledtabular}
\end{table}
\begin{table}
\caption{\label{tab:F-T-alpha} The time dependence of the third-order
perturbative solutions in the transverse modes
(the case of $\gamma=4/3$ in the E-dS universe model).
The contribution means the time component in the source terms.
The indices of the time evolution $\alpha$ 
(Eq.~(\ref{eqn:F-alpha})) depend
on the convolution in the source terms $Q^{L (3)}$.
Here we define 
$b_1 \equiv b(\bm{K}''), b_2 \equiv b(\bm{K}'-\bm{K}''), 
 b_3 \equiv b(\bm{K}-\bm{K}'), b_4 \equiv b(\bm{K}-\bm{K}'')$.
``P'' denotes that the contribution does not appear
in the dust model.}
\begin{ruledtabular}
\begin{tabular}{lllc}
The convolution & The contribution & The indices $\alpha$ &\\ \hline
$S^{(1)} (\bm{K}'') \ast S^{(2)} (\bm{K}-\bm{K}'')$
& $D E^L_1$ & $-\frac{1}{2} \pm b_1 \pm b_2
\pm b_3$ & \\
& $D E^L_2$ & $-\frac{1}{2} \pm \frac{1}{6} \pm b_1
\pm b_4$ & P \\
& $D E^L_3$ & $-\frac{1}{2} \pm \frac{1}{6} \pm
 \frac{1}{6} \pm b_1$ & \\
$S^{(1)} (\bm{K}'') \ast s^{T (2)} (\bm{K}-\bm{K}'')$
& $D E^T_1$ & $-\frac{1}{2} \pm b_1 \pm b_2
\pm b_3$ & P \\
& $D E^T_2$ & $-\frac{1}{2} \pm \frac{1}{6} \pm b_1
\pm b_4$ & \\
$s^{T (1)} (\bm{K}'') \ast S^{(2)} (\bm{K}-\bm{K}'')$
& $D^T E^L_1$ & $-\frac{1}{2} \pm \frac{1}{6} \pm b_2
\pm b_3$ & \\
& $D^T E^L_2$ & $-\frac{1}{2} \pm \frac{1}{6} \pm \frac{1}{6}
\pm b_4$ & P \\
& $D^T E^L_3$ & $-\frac{1}{2} \pm \frac{1}{6} \pm \frac{1}{6}
\pm \frac{1}{6}$ & \\
$s^{T (1)} (\bm{K}'') \ast s^{T (2)} (\bm{K}-\bm{K}'')$
& $D^T E^T_1$ & $-\frac{1}{2} \pm \frac{1}{6} \pm b_2
\pm b_3$ & P \\
& $D^T E^T_2$ &  $-\frac{1}{2} \pm \frac{1}{6} \pm \frac{1}{6}
\pm b_4$ & \\
$S^{(1)} (\bm{K}'') \ast S^{(1)} (\bm{K}'-\bm{K}'')
 \ast S^{(1)} (\bm{K}-\bm{K}')$
& $D D D$ & $-\frac{1}{2} \pm b_1 \pm b_2
\pm b_3$ & P \\
$s^{T (1)} (\bm{K}'') \ast S^{(1)} (\bm{K}'-\bm{K}'')
 \ast S^{(1)} (\bm{K}-\bm{K}')$
& $D^T D D$ & $-\frac{1}{2} \pm \frac{1}{6} \pm b_2
\pm b_3$ & P \\
$s^{T (1)} (\bm{K}'') \ast s^{T (1)} (\bm{K}'-\bm{K}'')
 \ast S^{(1)} (\bm{K}-\bm{K}')$
& $D^T D^T D$ & $-\frac{1}{2} \pm \frac{1}{6} \pm \frac{1}{6}
\pm b_3$ & P
\end{tabular}
\end{ruledtabular}
\end{table}

Tables~\ref{tab:F-L-alpha} and \ref{tab:F-T-alpha}
show the time dependence of the third-order
perturbative solutions. When the dust model
\cite{Sasaki98} and the pressure model are compared,
we notice that several
terms in the third-order perturbative solutions do not appear
in the dust model. For example, the contribution of the term
$D E^T_1$, which is proportional to $D E^L_1$, i.e., one of
most dominant term, appears only in the pressure model.

\section{Summary and Concluding Remarks}
\label{sec:summary}

In this paper, we showed the third-order Lagrangian perturbative equations
for the cosmological fluid with pressure. Then we derived third-order
perturbative solutions for a simple case.

In the pressure model, we must notice one other point. The longitudinal
mode in the first-order perturbation depends on the scale (or the
wavenumber) of the fluctuation. On the other hand, the transverse
mode in the first-order perturbation is independent of the scale
of the fluctuation. In the dust model, because the transverse
mode does not have growing modes, even if the primordial vorticity
exists, we can simply ignore this mode.
However in the pressure model, we cannot ignore it. The growing factor
in the longitudinal mode depends on the scale of the fluctuation;
when we consider small-scale fluctuation, the transverse mode
dominates.

The transverse modes do not appear in spacial one-dimensional model.
In two- or three-dimensional model, the computation of the
perturbation becomes huge. Especially, in the third-order
perturbation, enormous mode-coupling computation is required.
Even if in one-dimensional model~\cite{Tatekawa05A}, the computation
of the third-order approximation required long time.
When we compute the third-order approximation in generic case,
we will apply Monte Carlo integration~\cite{MonteCarlo} for
the computation of the mode-coupling.

Here we estimate importance of the transverse mode in the pressure
model. For simplicity, we consider only the E-dS Universe model.
In past papers~\cite{Morita01, Tatekawa02}, we showed that when
the polytropic index $\gamma$ is less than $4/3$, all
longitudinal modes oscillate and decay. On the other hand, when
the polytropic index $\gamma$ is more than $4/3$, all
longitudinal modes grow as those in the dust model do.
The case of $\gamma=4/3$ is critical. We consider the index of
the time component in the first-order perturbation
 (Eq.~(\ref{eqn:1st-L-sol})). If the Lagrangian wavenumber $\bm{K}$
satisfies the condition
\begin{equation}
\frac{25}{36} - B|\bm{K}|^2 > \frac{1}{36} \,,
\end{equation}
i.e.,
\begin{equation}
|\bm{K}| < \sqrt{\frac{2}{3B}} = \frac{2 \sqrt{6}}{5} K_J
\simeq 0.98 K_J \,,
\end{equation}
the longitudinal modes dominate during evolution. Therefore when
we consider small-scale ($|\bm{K}| > (2 \sqrt{6}/5) K_J$)
fluctuation with primordial vorticity, we must consider the evolution
of the transverse modes.

Recently several dark matter models have been proposed~\cite{DarkMatter}.
In several model, special interaction affects avoidance of matter concentration. In general, these interaction is introduced
by scattering cross-section. Instead of the cross-section,
if the interaction in some kind of
dark matter can be described by the effective pressure, we can examine
the behavior of the density fluctuation in a quasi-nonlinear stage.
Furthermore, when we compare the observations and the structure
that is formed by using the pressure model, we can delimit
the nature of the dark matter.
In future, when we consider the evolution of the density fluctuation
or the peculiar velocity in even high-density regions, although 
these seem highly complicated, the
third-order perturbative solutions may become useful.

\begin{acknowledgments}
We are grateful to Kei-ichi Maeda for his continuous encouragement.
We would like to thank Peter Musolf for checking of
English writing of this paper. 
This work was supported by the Grant-in-Aid for Scientific
Research Fund of the Ministry of Education, Culture, Sports, Science
and Technology (Young Scientists (B) 16740152).
\end{acknowledgments}

\end{document}